\documentclass[a4paper,11pt]{article}
\usepackage{pos}
\usepackage{bm}

\title{Composition of the inclusive semi-leptonic decay of $B$ meson}

\author[a]{Gabriela Bailas}
\author*[b,c]{Shoji Hashimoto}

\affiliation[a]{Center for Artificial Intelligence Research,
  University of Tsukuba, Tsukuba, Ibaraki 305-8577, Japan}
\affiliation[b]{Theory Center, Institute of Particle and Nuclear
  Studies, High Energy Accelerator Research Organization (KEK),
  Tsukuba, Ibaraki 305-0801, Japan}
\affiliation[c]{School of High Energy Accelerator Science, The
  Graduate University for Advanced Studies (SOKENDAI), 
  Tsukuba, Ibaraki 305-0801, Japan}

\emailAdd{shoji.hashimoto@kek.jp}

\abstract{
  Utilizing the approach recently proposed for the inclusive
  semi-leptonic decay rate on the lattice, we compute the differential
  decay rate of a $B_s$ meson for various kinematical channels.
  The results are compared with the contributions from the ground
  states ($D$ and $D^*$) as well as from the orbitally excited states
  ($D^{**}$'s).
  The computation so far is carried out with an unphysically light
  bottom quark and strange spectator quark.
}

\FullConference{%
 The 38th International Symposium on Lattice Field Theory, LATTICE2021
  26th-30th July, 2021
  Zoom/Gather@Massachusetts Institute of Technology
}

\begin{document}
\begin{flushright}
  KEK-CP-0386
\end{flushright}

\maketitle

\section{Introduction}
Semi-leptonic decays of $B$ meson have been measured in various
specific final states, such as $D^{(*)}\ell\nu$ or $D^{**}\ell\nu$.
They can be used to determine the Cabibbo-Kobayashi-Maskawa (CKM)
matrix element $|V_{cb}|$.
(Here, we focus on the $b\to c$ decays, but the formulation and
calculation method are applicable also for the $b\to u$ channels.)
On the other hand, the experimentalists can also perform the so-called
inclusive analysis, {\it i.e.} all possible final states including a
charm quark are counted.
Theoretically, such experimental results may be compared with the OPE
analysis \cite{Blok:1993va,Manohar:1993qn} to determine $|V_{cb}|$.

More recently, one of the authors proposed a method to compute the inclusive
decay rate using lattice QCD \cite{Gambino:2020crt}.
It utilizes a method to implicitly sum over all possible states
to appear in two operator insertions on the lattice
\cite{Hashimoto:2017wqo,Bailas:2020qmv}.
To be explicit, we consider the forward-Compton amplitude of the form
\begin{equation}
  \langle B(\bm{0})|\tilde{J}_\mu^\dagger(-\bm{q};t)
  \;\; \Big\vert \;\;
  \tilde{J}_\nu(\bm{q};0)|B(\bm{0})\rangle,
  \label{eq:compton}
\end{equation}
where two flavor-changing currents $J_\mu$ are inserted with a specific spatial
momentum $\bm{q}$.
The vertical line in the middle represents all possible states of the
specified quantum number that can contribute.
The key idea is to use the time separation $t$ to control the weight
among the states having different energies.
The method allows a fully non-perturbative computation of the
inclusive decay rate, so that one can verify the OPE-based calculation
that has been used so far.

There is a long-standing tension between the $|V_{cb}|$ determinations
from the exclusive and inclusive channels.
The aim of our study is to understand the cause of the problem by
providing a framework of theoretical calculation that can be applied
for both analyses.
We may also identify the contribution from the excited state $D$
mesons and $D\pi$ continuum states from the lattice data, which
would enable another test of the calculation by comparing with the
corresponding experimental data.

This contribution mainly describes the lattice computatin, and
the comparison with the OPE calculation for the same set of parameters
is separately presented in \cite{Gambino:2021zrp}.

\section{Inclusive decay rate: outline of the formalism}
Here we outline the method proposed in \cite{Gambino:2020crt}.

The differential decay rate of the $B$ meson can be decomposed as
$d\Gamma\sim |V_{cb}|^2 l^{\mu\nu}W_{\mu\nu} dq_0 d\bm{q}$,
where the leptonic tensor $l^{\mu\nu}$ is determined by the kinematics
of the final state leptons $\ell$ and $\nu$.
The momentum transfer to the lepton pair is $q=(q_0,\bm{q})$.
The hadronic tensor $W_{\mu\nu}$, on the other hand, has a complicated
structure
\begin{equation}
  W_{\mu\nu}(q_0,\bm{q}) \sim
  \sum_X(2\pi)^3\delta^{(4)}(p_B-q-p_X)
  \frac{1}{2m_B}
  \langle B(\bm{p}_B)|J_\mu^\dagger(0)|X\rangle
  \langle X|J_\nu(0)|B(\bm{p}_B)\rangle,
\end{equation}
where the flavor-changing current $J_\mu$ induces the decay $b\to c$,
and the sum of the state $X$ runs over all possible final states
including a charm quark.
When the initial $B$ meson is at rest, $\bm{p}_B=\bm{0}$, the final
hadronic state $X$ has a momentum $\bm{p}_X=-\bm{q}$ given by the
momentum transfer $q$.
Since it depends on the structure of the initial $B$ meson state, the
hadronic tensor $W_{\mu\nu}$ is also called the structure function.

We notice that the sum over the states $X$ may be considered as an
integral over its energy $p_X^0$.
Because of the $\delta$-function, it is given by $p_X^0\equiv\omega =
m_B-q_0$.
Thus, the structure function picks up the state of energy $\omega$,
and
\begin{equation}
  W_{\mu\nu}(\omega,\bm{q}) \sim \langle B(\bm{0})|
  \tilde{J}_\mu^\dagger(-\bm{q}) \delta(\omega-\hat{H})
  \tilde{J}_\nu(\bm{q}) |B(\bm{0})\rangle.
\end{equation}
Here $\tilde{J}_\mu(\bm{q})$ is a Fourier-transform of the current.
We specify the energy $\omega$ by inserting the $\delta$-function
$\delta(\omega-\hat{H})$ between the currents.
($\hat{H}$ is the Hamiltonian of QCD.)

The total decay rate can then be written as
\begin{equation}
  \Gamma\propto |V_{cb}|^2
  \int_0^{\bm{q}^2_{\mathrm{max}}}d\bm{q}^2
  \int_{\sqrt{m_D^2+\bm{q}^2}}^{m_B-\sqrt{\bm{q}^2}}d\omega
  \,
  K(\omega;\bm{q}^2)
  \langle B(\bm{0})|
  \tilde{J}_\mu^\dagger(-\bm{q}) \delta(\omega-\hat{H})
  \tilde{J}_\nu(\bm{q}) |B(\bm{0})\rangle,
  \label{eq:total}
\end{equation}
where $K(\omega;\bm{q}^2)$ is a kinematical factor originating from
the leptonic tensor.

The problem is then how to compute the integral over the
final-state energy $\omega$ with the weight of $K(\omega;\bm{q})$.
The upper limit of the $\omega$-integral is also imposed
by the kinematics; we may include them in $K(\omega,\bm{q}^2)$
using the Heaviside function as $\theta(m_B-\sqrt{\bm{q}^2}-\omega)$
while extending the upper limit to infinity.
Then, we can rewrite the $\omega$-integral using
\begin{equation}
  \int_0^\infty\!d\omega\, K(\omega;\bm{q}^2)
  \langle B(\bm{0})|
  \tilde{J}_\mu^\dagger(-\bm{q}) \delta(\omega-\hat{H})
  \tilde{J}_\nu(\bm{q}) |B(\bm{0})\rangle
  =
  \langle B(\bm{0})|
  \tilde{J}_\mu^\dagger(-\bm{q}) K(\hat{H};\bm{q}^2)
  \tilde{J}_\nu(\bm{q}) |B(\bm{0})\rangle.
  \label{eq:target}
\end{equation}
Here, the kinematical factor is promoted to an operator by replacing
$\omega$ by $\hat{H}$.
On the other hand, what one can calculate on the lattice is the
Compton amplitude (\ref{eq:compton}), which is also written in the
form
\begin{equation}
  \langle B(\bm{0})|
  \tilde{J}_\mu^\dagger(-\bm{q};t) 
  \tilde{J}_\nu(\bm{q};0) |B(\bm{0})\rangle
  =
  \langle B(\bm{0})|
  \tilde{J}_\mu^\dagger(-\bm{q};t) e^{-\hat{H}t}
  \tilde{J}_\nu(\bm{q};0) |B(\bm{0})\rangle,
  \label{eq:lat_compton}
\end{equation}
because the time separation between the currents may be described by
the transfer matrix $\exp(-\hat{H}t)$.

Given the similarity between (\ref{eq:target}) and
(\ref{eq:lat_compton}), one notices that (\ref{eq:target}) can be
evaluated if the kernel operator $K(\hat{H})$ can be approximated in
the form
$K(\hat{H})\simeq
k_0+k_1e^{-\hat{H}}+k_2e^{-2\hat{H}}+\cdots+k_Ne^{-N\hat{H}}$
with some coefficients $k_j$.
Such approximation can be constructed using the Chebyshev polynomials:
\begin{equation}
  K(\omega)\simeq\frac{c_0}{2}+\sum_{j=1}^N c_j^* T_j^*(e^{-\omega}),
\end{equation}
where $T_j^*(x)$'s are shifted Chebyshev polynomials defined as
$T_j^*(x)=T_j(2x-1)$ from the standard Chebyshev polynomial $T_j(z)$.
Thus, the shifted Chebyshev polynomials are defined for $0\le x\le 1$
or $\infty\ge\omega\ge 1$ since $x=e^{-\omega}$.
The coefficients $c_j^*$ can be easily computed for arbitrary kernel
operator $K(\omega)$. 
Since the Chebyshev polynomials $T_j^*(e^{-\omega})$ are
constructed from polynomials $(e^{-\omega})^m$ with positive integer
$m$'s, they are related to the transfer matrix $e^{-\hat{H}t}$
appearing in (\ref{eq:lat_compton}).
The forward Compton amplitude (\ref{eq:lat_compton}) can therefore be
used to approximate the target quantity (\ref{eq:target}).

The kernel function $K(\omega;\bm{q}^2)$ has the following structure 
\begin{equation}
  K(\omega;\bm{q}^2)\sim e^{2\omega t_0} (m_B-\omega)^l
  \theta(m_B-\sqrt{\bm{q}^2}-\omega).
\end{equation}
Here, $(m_B-\omega)^l$ ($l=$ 0, 1 or 2) originates from the leptonic
tensor; the factor $e^{2\omega t_0}$ is introduced to avoid any
divergence due to a contact term between the two currents by
normalizing the matrix element by the value at a small time separation $t_0$. 
The Heaviside $\theta$-function $\theta(m_B-\sqrt{\bm{q}^2}-\omega)$
implements the upper limit of the $\omega$ integral (see (\ref{eq:total})).

The Chebyshev approximation of the kernel function is harder when the
function contains a discontinuous (or even rapid) change such as that given by
the $\theta$-function, and we introduce a smearing to modify the
$\theta$-function to a smooth function with a smearing width $\sigma$.
(To be explicit we use the sigmoid function, but the details are not
important.)
We need to take the limit $\sigma\to 0$ to obtain the final result.

\begin{figure}[tbp]
  \centering
  \includegraphics[width=4.9cm]{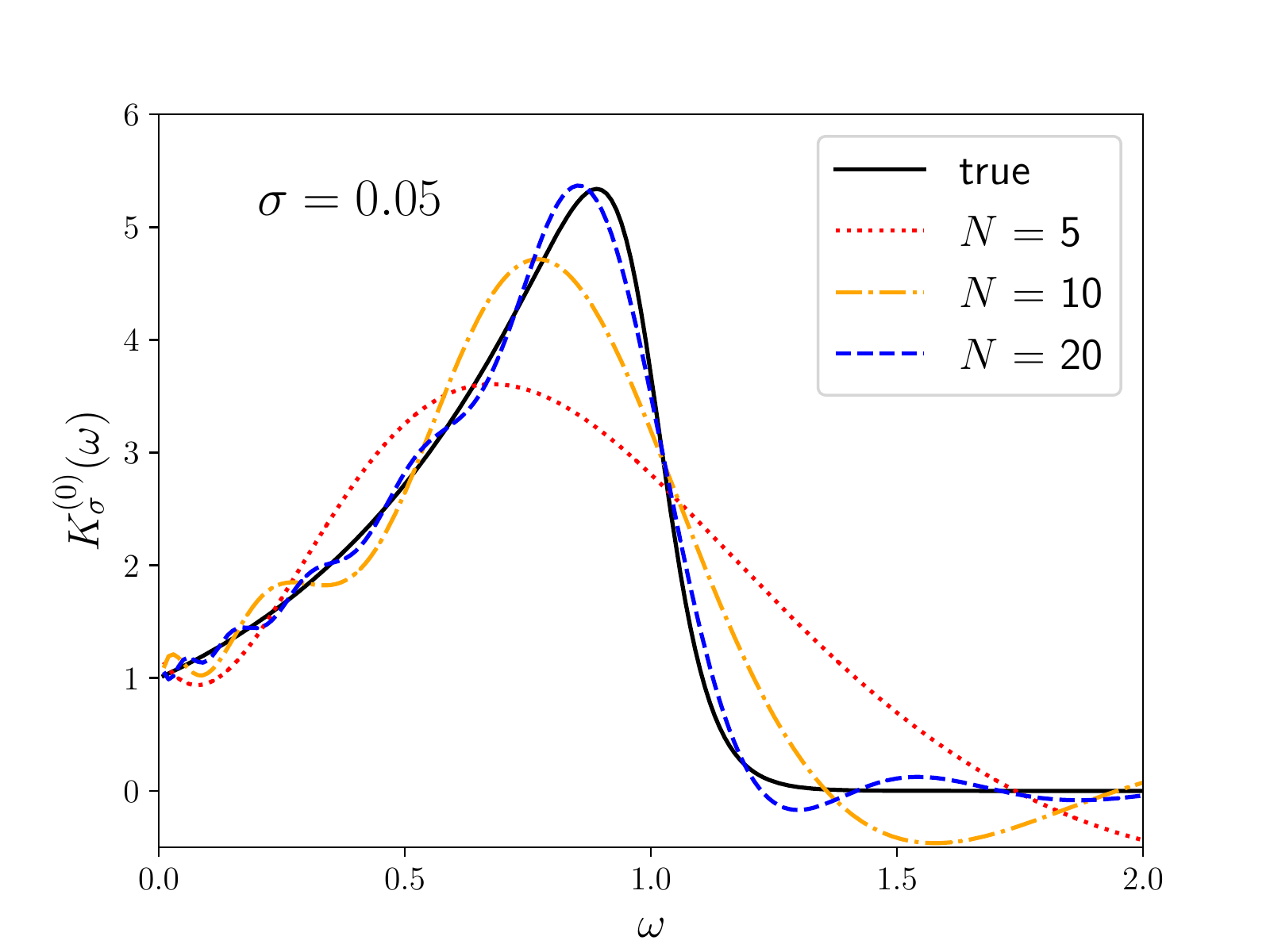}
  \includegraphics[width=4.9cm]{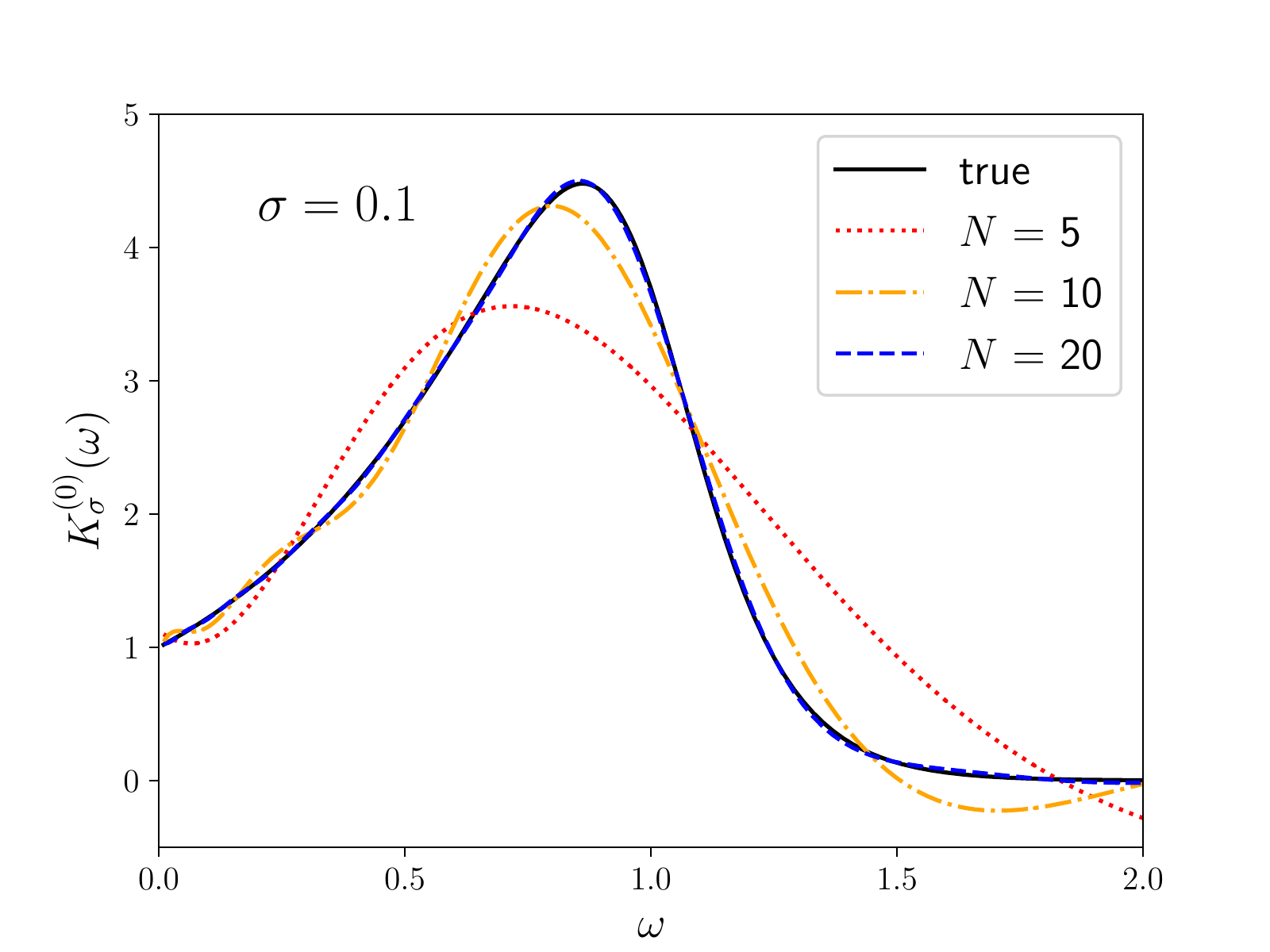}
  \includegraphics[width=4.9cm]{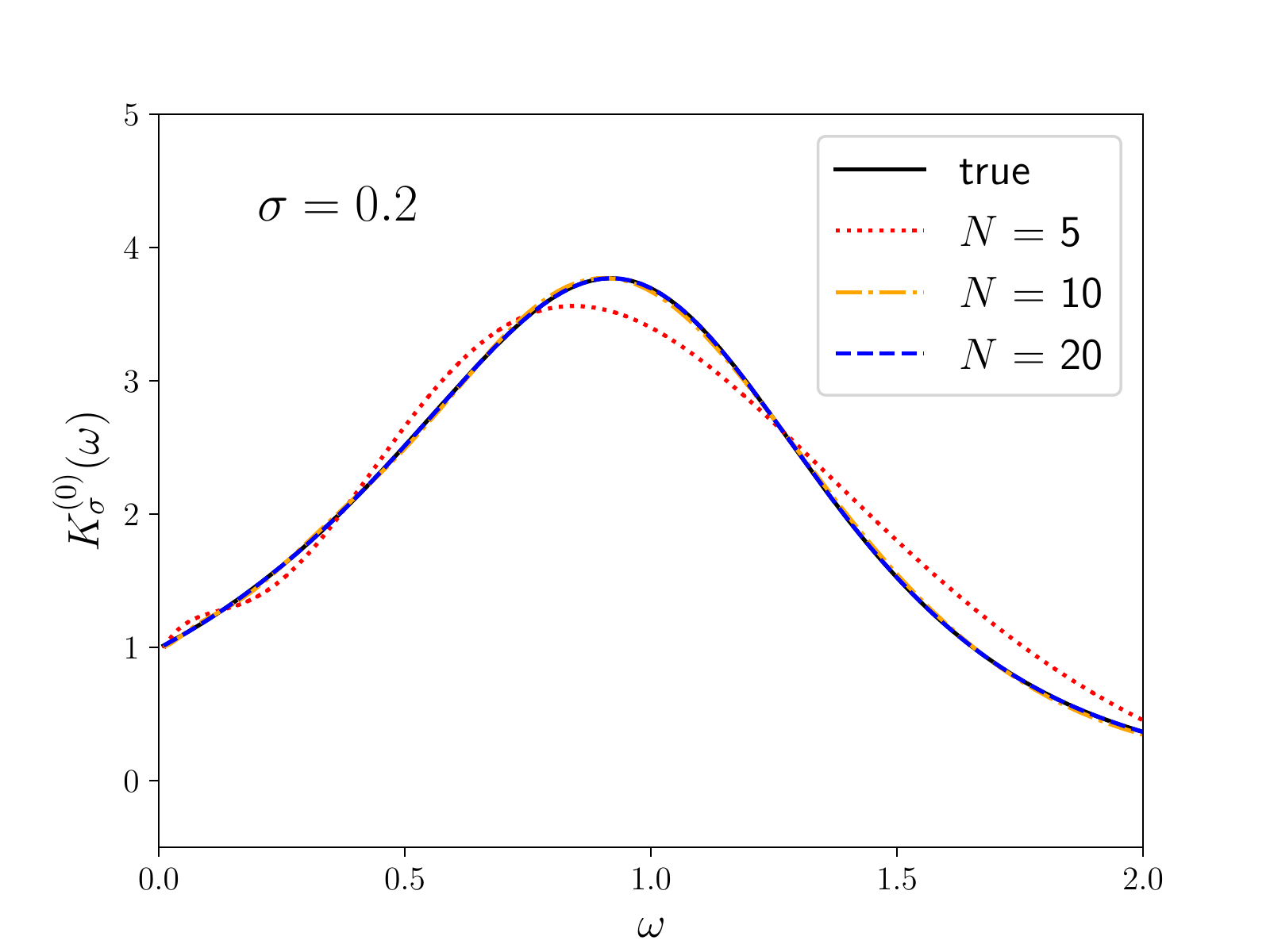}
  \caption{Kernel function for $l=0$ with smearing with the width
    $\sigma=0.05$ (left panel) 0.1 (middle) and 0.2 (right).
    Together with the true function (thich curve), Chebyshev
    approximations of order $N$ = 5, 10 and 20 are shown (dashed and
    dotted curves).
    The threshold is at $\omega\simeq$ 1.0.
    (All in the lattice unit.)
  }
  \label{fig:kernel}
\end{figure}

The kernel function $K(\omega)$ is plotted for $l=0$ with the smearing
width $\sigma$ = 0.05, 0.1 and 0.2 in Fig.~\ref{fig:kernel}.
Around the threshold of the $\theta$-function, the shape is smeared so
that the function is smoothed.
The Chebyshev approximations are also shown in the plots for the
order of polynomial $N$ = 5, 10 and 20.
The approximation is rather precise when the width is large ($\sigma$
= 0.2 on the right panel) even with the limited order of polynomials,
while it gets harder for small $\sigma$.
Although it would depend on the precision one wants to achieve, it
seems that at least $N$ = 20 is necessary to achieve a sensible
approximation for $\sigma$ = 0.05.

\section{Compton amplitude}
We compute the Compton amplitude of the form (\ref{eq:lat_compton}) on
the lattice.
It is obtained from a four-point function with two flavor-changing
$V-A$ currents inserted with time separation $t$ between interpolating
operators to create or annihilate the $B$ meson.
To compute the amplitude with all possible combinations of spin
orientations $\mu$ and $\nu$ we have to repeat the computation of
sequential sources many times, as well as for the choices of final
momenta. (The initial $B$ meson is set at rest.)

The computation is done on a lattice of $48^3\times 96$ at $1/a\simeq$
3.6~GeV generated including up, down and strange sea quarks described
by the Mobius domain-wall fermions.
The ensemble is the same as that used in \cite{Gambino:2020crt} and is
a part of the large set of ensembles generated and used in
\cite{Nakayama:2016atf}.
The valence quarks are also domain-wall fermions; the charm quark mass
is tuned to its physical value while the bottom quark mass is taken
(unphysically) light and about 2.7~GeV.
The spectator quark is strange in this study, so that the initial
state is (unphysically light) $B_s$ meson.

The statistics is 100 gauge configurations and the measurement is
repeated four times on each configuration with different source time
slices. 

\begin{figure}[tbp]
  \centering
  \includegraphics[width=10cm]{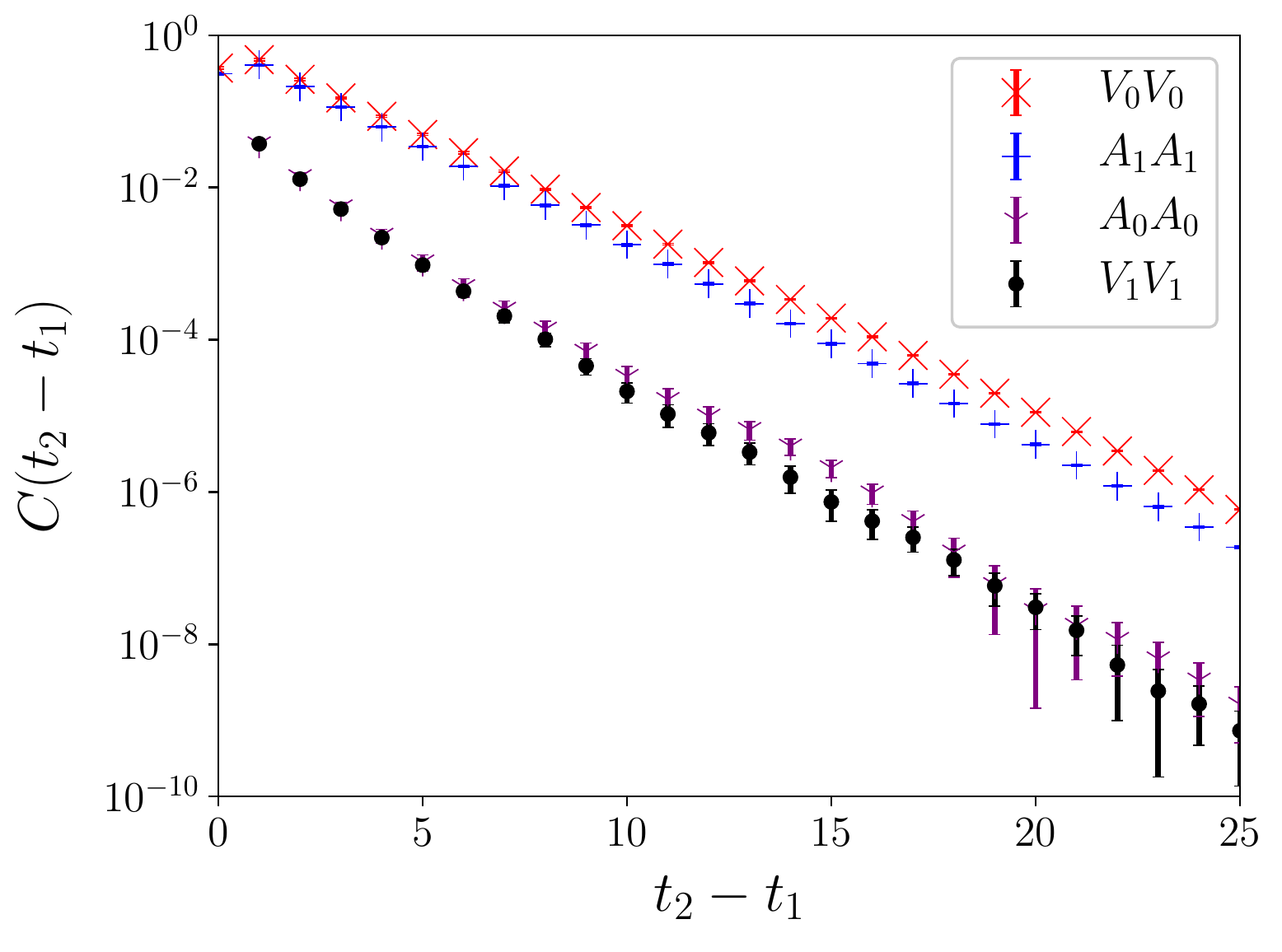}
  \caption{Compton amplitude as a function of the time separation.
    The final state has a vanishing spatial momentum.
    Different current insertions are shown together: vector ($V$) and
    axial-vector ($A$) currents in the temporal ($0$) and spatial
    ($k=1$) directions.
  }
  \label{fig:corr}
\end{figure}

Fig.~\ref{fig:corr} shows the Compton amplitude
(\ref{eq:lat_compton}) as a function of the time separation $t\equiv
t_2-t_1$ for the case of zero spatial momentum in the final state.
The current insertions are either $V(t_2)V(t_1)$ or $A(t_2)A(t_1)$
combinations of vector ($V$) or axial-vector ($A$) currents.
Non-zero values are obtained for $(\mu,\nu)=(0,0)$ (temporal) or
$(\mu,\nu)=(k,k)$ (spatial) orientations of the currents.
Among them, the $V_0V_0$ and $A_kA_k$ combinations show the largest
signal due to the $S$-wave ground-state contributions from the
pseudo-scalar $D_s$ ($0^-$) or vector $D_s^*$ ($1^-$) meson.
If we look into the details, the amplitude of $V_0V_0$ is slightly
larger than $A_kA_k$ because the pseudo-scalar state is lighter.

The other channels $A_0A_0$ and $V_kV_k$ show substantially
($\times 50$) smaller contributions.
They correspond to the scalar ($0^+$) and axial-vector ($1^+$) states,
respectively.
In the quark model, they are $P$-wave states and their coupling to the
initial $B_s$ meson is weak.
According to the analysis based on the heavy quark effective theory
(HQET) \cite{Leibovich:1997em}, the $B\to D^{**}\ell\nu$ ($D^{**}$
denotes the $P$-wave states generically) form factor is suppressed as
$\bar{\Lambda}/2m_c$ with $\bar{\Lambda}$ a typical QCD scale $\sim$
300~MeV compared to the $B\to D^{(*)}\ell\nu$ form factors, which is
$O(1)$, so that the Compton amplitude is suppressed by
$(\bar{\Lambda}/2m_c)^2\sim$ 0.01.

These opposite parity channels can also couple to $DK$ continuum,
which is heavier than the above mentioned $P$-wave mesons.
Interestingly, the data show the contributions of such excited states
at small time separations $t_2-t_1\lesssim$ 5.

\section{Inclusive decay rate}
According to (\ref{eq:total}), we obtain the total decay rate from the
Compton amplitudes as described in the previous section.
The kinematical factor $K(\omega;\bm{q}^2)$ is promoted to the
integral kernel for the $\omega$-integral.
The spatial momentum integration over $\bm{q}^2$ is yet to be performed.

\begin{figure}[tbp]
  \centering
  \includegraphics[width=10cm]{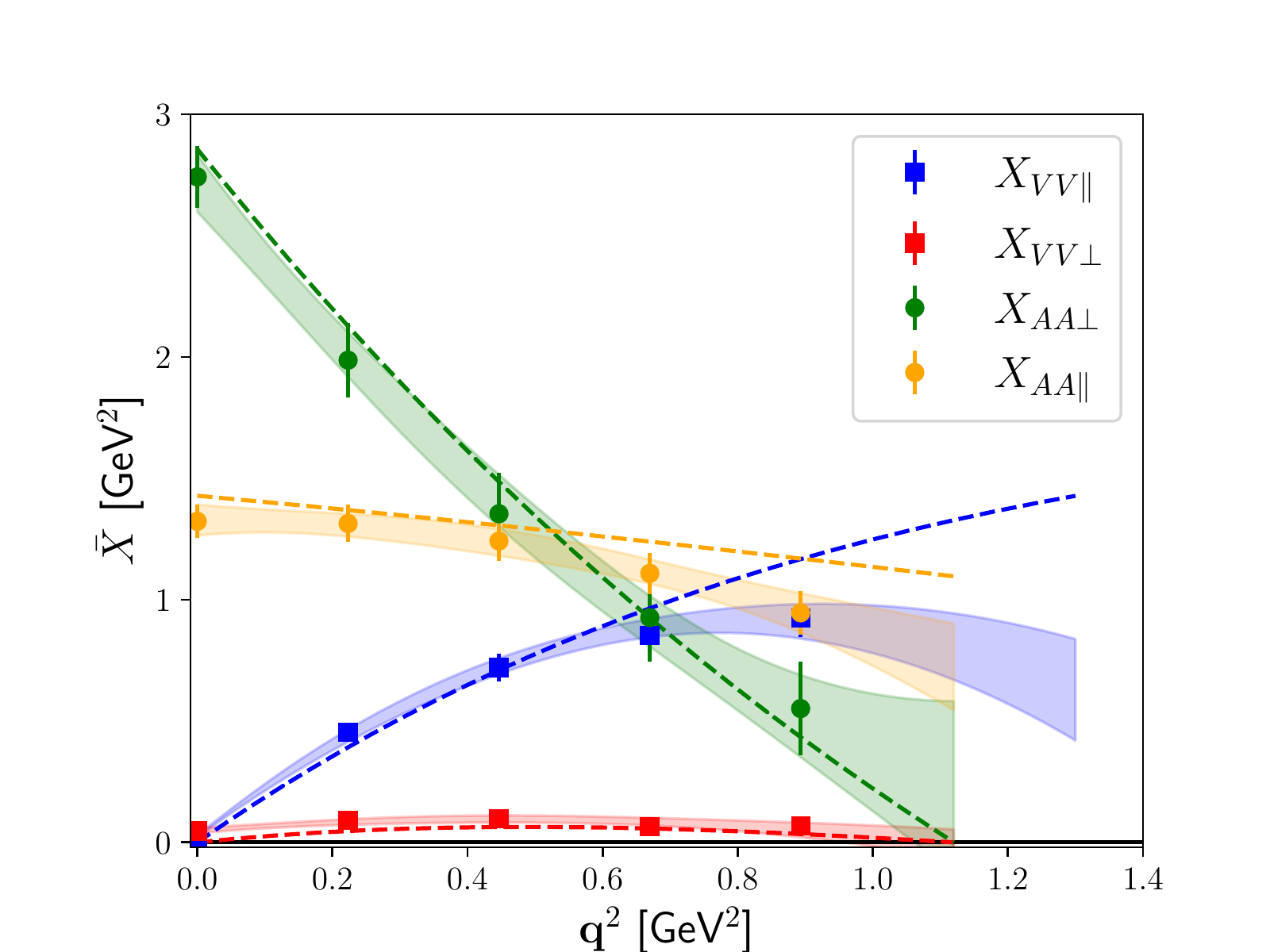}
  \caption{
    Differential decay rate divided by $|\bm{q}|$ as a function
    of $\bm{q}^2$.
    The results are shown for four different channels: $VV_\parallel$,
    $VV_\perp$, $AA_\parallel$ and $AA_\perp$.
    (For more details, see the text.)
    The bands are contributions of ground state, $D$ or $D^*$ meson,
    as estimated using the corresponding form factors.
  }
  \label{fig:diff}
\end{figure}

The results for the differential decay rate (divided by $|\bm{q}|$)
are plotted in Fig.~\ref{fig:diff} as a function of $\bm{q}^2$.
Four contributions are plotted separately:
$VV$ and $AA$ denote the two currents inserted, which are decomposed
from the $V-A$ weak currents.
The other possible combinations, $VA$ and $AV$, do not contribute for
the differential decay rate after integrating over the lepton energy
$E_\ell$.
The current orientations, $\parallel$ and $\perp$, distinguish the
currents in the direction parallel or perpendicular to the momentum
$\bm{q}$.
For instance, when the momentum $\bm{q}$ is in the $z$-direction,
$(0,0,1)$, the $VV\!\perp$ represent the contribution from $V^1V^1$ and
$V^2V^2$, while $VV\!\parallel$ is that of $V^3V^3$, $V^3V^0$, $V^0V^3$ and $V^0V^0$.

Also shown by the bands are the corresponding contributions from the
ground states, {\it i.e.} from
$B_s\to D_s\ell\nu$ for $VV\!\parallel$ and from $B_s\to D_s^*\ell\nu$ for
others.
We have separately computed the form factors of these decay modes,
$h_+(w)$, $h_-(w)$ and $h_{A_1}(w)$, $h_{A_2}(w)$, $h_{A_3}(w)$,
$h_V(w)$.
(The computation is done along with \cite{Kaneko_lat21}, and the
analysis method is essentially the same.)
Using these results, the differential decay rate for each channel can be
constructed.

As one can see, the inclusive decay rate is dominated by the ground
state contributions for each channel.
This is not unreasonable because the initial state is lighter
than the physical $B_s$ meson.
Kinematically there is not so much room to produce extra particles other
than $D_s^{(*)}$.
Also, the heavy quark symmetry implies that the wave function of light
degrees of freedom are very similar between the initial $B_s$ and
final $D_s^{(*)}$ states, so that the overlap between these states are
enhanced when the initial and final quark masses are similar,
especially near the zero-recoil ($\bm{q}\sim 0$) limit.

As we mentioned in the previous section, the contribution from the
excited states including those from the $P$-wave meson is visible in
the Compton amplitudes.
It is however not significant for the differential decay rate, since
their contributions are roughly 50 times smaller.

\section{Summary}
Framework to compute inclusive decay rate on the lattice is now
available.
The essential step is the reconstruction of the energy integral from
Euclidean lattice correlators corresponding to the Compton amplitudes.
In our case, it is implemented using the Chebyshev approximation.
Other methods, such as the Backus-Gilbert method can also do the job
\cite{Hansen:2017mnd,Hansen:2019idp}.
The method to compute the energy integral can be also applied for
calculations of (not so) deep inelastic scattering cross section,
$\sigma(\ell N\to \ell' X)$ \cite{Fukaya:2020wpp} as well as those of
spectral sum of hadronic vacuum polarization \cite{Ishikawa:2021txe}.

The Compton amplitudes contain the contributions from ground states
and excited states.
With the lattice parameters taken in this work the ground state
contributions dominate the differential cross section, although
the excited state contributions such as those from the $P$-wave states
are visible in the Compton amplitude.

The present results can already been used to compare with the analytic
OPE calculations.
Even though the $b$-quark mass is not tuned to the physical value, the
comparison would still yield useful tests of the theoretical
approaches \cite{Gambino:2021zrp}.

\section*{Acknowledgement}
This work is a part of the efforts to make a comparison between
lattice and OPE calculations of inclusive semileptonic $B$ decays,
being done in collaboration with Paolo Gambino and Sandro Machler.
We thank the members of the JLQCD collaboration for helpful
discussions and providing the computational framework and lattice
data.
Takashi Kaneko, in particular, provided the unpublished data of
the $B_s\to D_s^{(*)}$ form factors computed on the same ensemble.
This work is supported in part by JSPS KAKENHI grant number 18H03710
and by the PostK and Fugaku supercomputer project through the Joint
Institute for Computational Fundamental Science (JICFuS).

\end{document}